\documentclass[12pt]{article}

\usepackage{latexsym,epsfig,colordvi}

\usepackage{scicite}

\usepackage{times}

\newcommand       \be           {\begin{equation}}
\newcommand       \ee           {\end{equation}}

\newcommand{\beq}{\begin{equation}}
\newcommand{\eeq}{\end{equation}}
\newcommand{\bea}{\begin{eqnarray}}
\newcommand{\eea}{\end{eqnarray}}
\newcommand\aj{AJ}%
\newcommand\apj{ApJ}%
\newcommand\apjl{ApJ}%
\newcommand\nat{Nature}%
\newcommand\mnras{MNRAS}%
\newcommand\aap{Astron. Astroph.}%
\newenvironment{sciabstract}{%
\begin{quote} \bf}
{\end{quote}}

\newcounter{lastnote}
\newenvironment{scilastnote}{%
\setcounter{lastnote}{\value{enumiv}}%
\addtocounter{lastnote}{+1}%
\begin{list}%
{\arabic{lastnote}.}
{\setlength{\leftmargin}{.22in}}
{\setlength{\labelsep}{.5em}}}
{\end{list}}

\title{The Use of Transit Timing to \\
Detect Extrasolar Planets \\
with Masses as Small as Earth}

\author
{Matthew J. Holman,$^{1\ast}$ Norman W. Murray$^{2,3}$\\
 \\
\normalsize{$^{1}$Harvard-Smithsonian Center for Astrophysics,}\\
\normalsize{MS51, 60 Garden Street, Cambridge, MA 02138, USA}\\
\normalsize{$^{2}$Canadian Institute for Theoretical Astrophysics,
University of Toronto,}\\
\normalsize{60 St. George Street, Toronto, ON M5S 3H8, Canada}\\
\normalsize{$^{3}$Canada Research Chair}\\
\\
\normalsize{$^\ast$To whom correspondence should be addressed; E-mail:
mholman@cfa.harvard.edu.}
}

\date{}

\begin{document} 




\maketitle 

\begin{center}
Submitted to {\it Science}, 17 November 2004.
\end{center}


\begin{sciabstract}
Future surveys for transiting extrasolar planets, including
the space-based mission Kepler~\cite{Borucki.2003}, are
expected to detect hundreds of Jovian mass planets and tens of
terrestrial mass planets.  For many of these newly discovered planets,
the intervals between successive transits will be measured with an
accuracy of 0.1--100 minutes. We show that these timing measurements
will allow for the detection of additional planets in the system (not
necessarily transiting), via their gravitational interaction with the
transiting planet.  The transit time variations depend on the mass of
the additional planet, and in some cases Earth-mass planets will
produce a measurable effect.

\end{sciabstract}

The one hundred or so currently known extrasolar
planets~\cite{Schneider} have revealed themselves through three types
of phenomena: 1) short-duration brightness anomalies in gravitational
microlensing events caused by planets near the lens
star~\cite{Bond.2004}, 2) reflex motions of the central star (as revealed by
radial velocity variations in the stellar spectrum or radio pulse arrival
times)~\cite{Wolszczan.1992,Backer.1993,Mayor.1995,Marcy.1996}, and 3)
variations in the apparent stellar brightness caused by a planetary
transit (passage of the planet in front of the
star)~\cite{Charbonneau.2000,Henry.2000,Udalski.2002a,Udalski.2002b,Udalski.2003}.
These approaches provide complementary information.  Gravitational
microlensing measurements primarily constrain the ratio of planet mass
to stellar mass.  Microlensing surveys are sensitive to
Earth-mass planets in principle but reveal little orbital information about the
planets discovered.  Radial velocity measurements allow for estimates
of the orbital period, eccentricity, and minimum mass of the planet.
With present technology, radial velocity surveys can only detect
planets with masses greater than about 10~Earth masses (orbiting low
mass stars)~\cite{McArthur.2004,Butler.2004,Narayan.2004}. Current
transit observations allow for the determination of the orbital period
and planetary radius.  Transit surveys, particularly if space-based,
will be sensitive to planets as small as Mercury's for the
smallest stars observed~\cite{Borucki.2003}.  Barring significant
improvements in the precision of radial velocity measurements, the
measurement of the mass and radius (and thus the density) of an
Earth-sized extrasolar planet would appear to be out of reach.

Here we point out that variations in the time interval between transits,
produced via gravitational interactions with additional planets, 
allow for the orbital period and mass of the
additional planet to be determined from transit observations alone.
In some instances, when two or more planets transit the same star, the
density of the planets can be determined.  This opens the
possibility of obtaining considerably more information about
transiting planets than has been previously thought.

The time interval between successive transits of an {\it unperturbed}
planet is always the same, because the orbital period is constant.
However, it has long been known that the presence of a third body can
produce short-term variations, in addition to the more familiar long-term variations, of the period
of a binary~\cite{Brown.1936,Soderhjelm.1975}. The same is true for
planetary systems.  It has been recently noted that the interval
between successive transits of the extrasolar planet HD~209458b would
vary by $\pm 3$~s if a second planet (of mass $10^{-4}~M_\odot$,
period 80-day, and eccentricity $e\sim 0.4$) existed in that
system~\cite{Bodenheimer.2003}.

Over the course of their orbits the transiting planet and a second planet
exchange energy and angular momentum as a result of their mutual gravitational
interaction.  This interaction, greatest at each planetary
conjunction, results in short-term oscillations of the semimajor axes
and eccentricities of the planets, which in turn alter the interval
between successive transits.  We first illustrate this effect by
considering our solar system.  Figure~1 shows the variation in the
transit intervals of our terrestrial planets, recorded by distant
observers located in the orbital plane of each planet.  The
gravitational perturbations among the planets in our solar system lead
to transit interval variations ranging from tens of seconds, for
Mercury, to thousands of seconds, for Mars.  The variations for Earth
and Venus show oscillations with the 583-day Earth-Venus synodic
period.

Next we investigate the influence of an additional planet on the
transit interval of HD~209458b.  We integrate the heliocentric
equations of motion of the hypothetical two-planet
system~\cite{Stoer.1980} and test that the relative energy error of
the system is less than $10^{-12}$ to assure the reliability of our
results. We assume that the two planets are coplanar, have orbits that
are perpendicular to the sky plane, and have initially aligned orbital
apsides.  We include the mutual gravitational interactions of the
planets and terms that account for the general relativistic (GR)
influence of the central star~\cite{Quinn.1991} but neglect the terms
for the GR influence of the planetary masses, terms for the oblateness
of the star, as well as terms for the tidal interaction with the star.
During the course of each integration, we iteratively solve for the
central transit times of the inner planet.

Figure~2 displays the interval between the times of successive
transits of HD~209458b, with each panel showing the results for a
different set of orbital parameters of a hypothetical exterior planet.
The large spikes in the transit interval, evident in the top three panels
of figure~2, occur near the times when the outer planet reaches its
periastron.  Hence, the orbital periods of both planets can be
determined directly.  We choose the period and eccentricity of the
four hypothetical planets
such that their pericenter distances are roughly the same.  As a result, the largest
transit interval variations have comparable magnitude.  For clarity, we
separate the light-time effect~\cite{Hilditch.2001} (due to the
varying distance between the star and the observer as the star moves
with respect to the center of mass among the star and planets) from
the dynamical effects described above.

Whether the presence of a given companion planet can be detected from
short-term transit interval variations alone depends on the difference
between the minimum and maximum transit interval.  The black lines in
figures~3~and~4 show those differences for Jupiter-mass
($10^{-3}~M_\odot$) and Earth-mass ($3\times10^{-6}~M_\odot$)
perturbing planets, respectively, as a function of their orbital
period (or semimajor axis) and eccentricity.  In these calculations
the star has mass $M_*= M_\odot$; the transiting planet has the mass
of Jupiter and an initial eccentricity and period of $e_1=0.01$ and
$P_1=3.0~\mathrm{d}$, respectively.  For each perturber eccentricity,
a range of perturber periods are tested, in increments of 0.1~d,
starting with the minimum perturber periods that ensures that the
orbits of the two planets do not initially cross.  We follow the same
numerical procedure described earlier.  During each integration,
simulating $10^4$~d, we record the times of transit, from which the
minimum and maximum intervals between transits were determined.  As
expected, for a given perturber eccentricity the period variation
decreases as the perturber period and semimajor axis increase.
Likewise, for a given perturber period, the variation is greater for
larger perturber eccentricity.  As indicated earlier, the transit
interval variation is primarily a function of the periastron distance
of the perturber.  Comparison of the two figures confirms that the
magnitude of the effect is proportional to the perturber mass.  We
also note that, for short time spans, the magnitude of the effect is
independent of the mass of the transiting planet (a result that
follows from the equivalence of inertial and gravitational mass).

Analytic estimates of the variation in transit intervals can be
found by integration of Lagrange's equations of planetary
motion~\cite{Murray.1999}.  Given a transiting planet with semimajor axis $a_1$ and period $P_1$,
and a perturbing planet with semimajor axis $a_2$ (assumed larger
than $a_1$), period $P_2$, and mass $M_2$, we find 
\begin{equation}
\Delta t \sim \frac{45\pi}{16} \left({M_2\over M_*}\right) P_1
\alpha_e^3 \left(1 - \sqrt{2}\alpha_e^{3/2}\right)^{-2} 
\end{equation}
where $\alpha_e = \frac{a_1}{a_2\left(1-e_2\right)}$.  The red dashed
lines in figures~3~and~4 show the estimate given by the expression
above.  Equation~(1) was derived by assuming the perturber follows a
parabolic orbit with a periastron distance of $a_2 (1-e_2)$.  It
significantly underestimates the actual variation in transit period
for small period ratios and is best suited for $e_2 \ge 0.3$.  The
transit interval variations increase with $P_1$ when the period ratio
of the two planets is held fixed.  Thus, the detection of companions
is easier for systems where the transiting planet is farther from the
star.  The timing variations are also larger for planets orbiting less
massive stars, for a given perturbing planet mass.

As suggested by figure~2, the eccentricity of the outer planet can be
estimated from the relative magnitudes of the variations $\Delta
t_{max}/\Delta t_{min}$.  If the period of the perturbing planet is
much greater that of the transiting planet, the final factor of
equation~(1) can be ignored.  The resulting equation can be rearranged to
provide an estimate of the mass of the perturbing planet
\be
M_2 = \frac{16}{45\pi} M_* \frac{\Delta t_{max}}{P_1}\left(\frac{P_2}{P_1}\right)^2
\left(1-e_2\right)^3,
\ee
given an estimate of $e_2$.

Large excursions in transit interval variation in figures~3~and~4 occur
near small integer ratios of the orbital periods of the two planets.
These correspond to mean-motion resonances, near which the planets
undergo larger oscillations of semimajor axis and
eccentricity~\cite{Murray.1999}.  The width in semimajor axis of each resonance region is
proportion to $a_2\left(M_*/M_2\right)^{1/2}$ and grows rapidly with
increasing eccentricity~\cite{Murray.1999}.  The ranges of perturber
periods in figures~3~and~4 in which the transit interval variations
appear irregular correspond to dynamical chaos resulting from the
overlap of adjacent mean-motion resonances~\cite{Wisdom.1980}.   We
have excluded the ranges of perturber period in which this 
chaos results in short-term dynamical instability (since such
planetary configurations are unlikely to be found).

The two planets of the GJ~876 system provide an excellent example of a
2:1 mean-motion resonance in an extrasolar planetary system.  Due to
the resonant gravitational interaction between the two planets, the
orbital periods of the inner and outer planet vary from 30.1~d to
31.1~d and 60.0~d to 59.1~d, respectively, over a libration period of
550~d~\cite{Laughlin.2001}.  Although careful photometric monitoring
has excluded the possibility of transits of the inner planet of the
GJ~876 system~\cite{transitsearch}, such transit interval variations would
be clearly seen if the inner planet transited.

In some cases, the libration period of a resonant system can be too
long for the transit interval variations due to the resonance to be
easily observed, even if the amplitude of the variation is large.
GJ~876's short libration period, $P_{lib}$, results from the
combination of its low-mass star ($M_*=0.4~M_\odot$) and massive
planets ($M_1 = 2~M_{jup}$, $M_2=4~M_{jup}$), since $P_{lib} \propto
P_1\left(M_*/M_2\right)^{1/2}$~\cite{Laughlin.2001,Murray.1999} (and
depends upon the order of the resonance).  Note that the $10^4$~d
integrations used for figures~(3)~and~(4) are long enough to sample a
full libration period for the low-order resonances.  A corresponding
system with a solar-mass star, a 1-year orbital period for the inner
planet, and jupiter-mass planets in the 2:1 resonance would have a
libration period of roughly 50~years.  For earth-mass planets the
libration period would be nearly 1000~years.  These long-term effects
would not be observable.  However, the smaller transit interval
variations that occur on the time scale of the orbital periods of the
two planets, that are superimposed on longer period variations, could
be observed.

The feasibility of this technique to detect additional planets in
transiting systems depends on the physical and orbital properties of
the multiple planet systems, as well as the accuracy with which the
times of transit can be measured.  Our expectations for systems with
multiple Neptune-mass to Jupiter-mass planets are guided by recent
discoveries, as well as by the giant planets in our solar system.  Fourteen (of 117) systems
are known to have two or more planets.  These have period ratios ranging
from 2 to 150, and three are known or thought to be in mean motion
resonances.  In the solar system, neighboring giant planets have
period ratios of two to three.  The eccentricities of
the planets in the solar system are small, but in many of the
extrasolar planetary systems the eccentricities are substantial,
ranging up to $e=0.7$.  

Both the theory of terrestrial planet formation and available
observations suggest that the typical terrestrial planet system will
have a configuration that results in variations in the transit
interval of hundreds of seconds.  Because the escape velocity from the
surface of a terrestrial planet is smaller than the escape velocity
from a solar mass star at roughly one AU, all the solid material in
the region from a few tenths to one or two AU must either accrete into
planets or fall onto the star.  The resulting planets are as closely
spaced as dynamical stability
permits~\cite{Chambers.1998,Kokubo.1998}. As observed in
our solar system and in the pulsar planet system, the period ratios
are of order two.  

Assuming the observations during the ingress and egress of a transit
are well-sampled, the error $\sigma_{t_c}$ (due to photon statistics alone)
associated with the measured time of the center of a transit of
duration $t_T$ is given by
\begin{equation}
\frac{\sigma_{t_c}}{t_T} \sim \left(\Gamma
t_T\right)^{-1/2}\rho^{-3/2},
\end{equation}
where $\Gamma$ is the photon count rate of the star and $\rho =
R_p/R_*$ is the ratio of the planet radius to the stellar radius.
Kepler~\cite{Borucki.2003}, a NASA Discovery mission, will monitor
100,000 A-M dwarf stars brighter than apparent magnitude $V=14$,
looking for transits of detecting Earth-sized planets in the
``habitable zone''~\cite{Kasting.1993}.  For Kepler, with its
$0.95~\mathrm{m}$ diameter aperture, $\Gamma = 7.8\times
10^8~10^{-0.4\left(V-14\right)}~\mathrm{hr}^{-1}$, for a star of
apparent magnitude $V$.  For a Jupiter-sized planet in a 1-year orbit
about a solar-mass star with $V=14$, we find
$\sigma_{t_c}\sim~20~\mathrm{s}$ ($\rho\sim0.1$, $t_T\sim13\mathrm{hr}$).  For an Earth-sized
planet, $\sigma_{t_c}\sim500~\mathrm{s}$ ($\rho\sim0.01$).  These
accuracies suggest that the transit period variations due to the
gravitational influence of Earth-mass planets with small period ratios can be detected by
Kepler.  For brighter stars observed with large-aperture ground-based telescopes,
the presence of more distant additional planets can be detected.  For
example, observing a $V=9$ star with a $6.5~\mathrm{m}$ aperture
telescope, $\sigma_{t_c}\sim~0.2~\mathrm{s}$.

One would like to measure the density of a transiting planet to determine
if it is a rocky terrestrial planet such as the Earth, an ice/water
giant such as Neptune, or a gas giant planet such as Jupiter. However,
when a transiting planet is perturbed by another planet that does not
transit, photometry yields an estimate of the radius of the transiting
planet but not its mass, and the transit timings yield an estimate of
the mass of the perturbing planet but not its radius.  In the cases in
which both planets transit their star the masses and radii of both
planets can be estimated, allowing density determinations.  This means
that for some subset of the discovered planets (double transit systems), radial velocity
observations will not be necessary to determine their masses and
densities.   This has important implications
for the search for habitable planets.

If transits of one planet are seen, what is the probability of seeing
transits of a second planet?  Assuming such a planet exists and that
the orbits of the two planets are coplanar, the probability that the
second planet also transits is simply $a_1/a_2$ if $a_2 > a_1$ (if
$a_2 < a_1$ transits of the second planet are assured in the case of coplanar
orbits).  If the orbits of the two planets are mutually inclined and
one planet transits the center of the star, the probability that the
second planet also transits is
\begin{equation}
P_{t_2} = \frac{2}{\pi}\arcsin\left(\frac{R_*}{a_2\sin i^\prime}\right),
\end{equation}
where $i^\prime$ is the mutual inclination between the orbits of the
two planets rather than the sky plane inclination. This assumes $\sin i^\prime >
R_*/a_2$, otherwise transits of the second planet are certain. For a solar-radius
star, a mutual inclination of a few degrees, and a semimajor axis
$a_2\sim 1~\mathrm{AU}$, the probability is roughly 10\%.  Thus, such
double transiting systems are likely to be found.

The Kepler team plans to search for transits with a consistent period,
depth, and duration~\cite{Borucki.2003}.  We caution that any
detection algorithm that looks for evidence of periodic transits in
the photometry must allow for variations in the transit period due to
the perturbations from unseen planets.  An
overly restrictive test for periodicity might reject some of the most
interesting and informative planetary systems.  The GJ876 system
discussed above, an extreme case, could easily be rejected as only
quasiperiodic.

We also caution that experiments looking for gradual transit period
variations due to slow secular trends must allow for the possibility
of significant short-term variations in the transit period.  Transit
period variations due to orbital precession induced by the general
relativistic effects of the central star, the oblateness of the star,
or the presence of a distant additional planet~\cite{Miralda.2002} or
due to the decay of the transiting planet's orbital semimajor axis as
a result of dissipative tidal interaction between the star and
planet~\cite{Sasselov.2003}, require many years to elapse before they
can be detected; without careful monitoring in the interim to
determine the short-term variations, the results of such experiments
could be easily misinterpreted.



\bibliographystyle{Science}

\begin{thebibliography}{10}

\bibitem{Borucki.2003}
W.~J. {Borucki}, {\it et~al.\/}, {\it Future EUV/UV and Visible Space
  Astrophysics Missions and Instrumentation. Edited by J. Chris Blades, Oswald
  H. W. Siegmund. Proceedings of the SPIE, Volume 4854, pp. 129-140 (2003).\/}
  (2003), pp. 129--140.

\bibitem{Schneider}
J.~{Schneider} (2004). {http://www.obspm.fr/encycl/catalog.html}.

\bibitem{Bond.2004}
I.~A. {Bond}, {\it et~al.\/}, {\it \apjl\/} {\bf 606}, L155 (2004).

\bibitem{Wolszczan.1992}
A.~{Wolszczan}, D.~A. {Frail}, {\it \nat\/} {\bf 355}, 145 (1992).

\bibitem{Backer.1993}
D.~C. {Backer}, R.~S. {Foster}, S.~{Sallmen}, {\it \nat\/} {\bf 365}, 817
  (1993).

\bibitem{Mayor.1995}
M.~{Mayor}, D.~{Queloz}, {\it \nat\/} {\bf 378}, 355 (1995).

\bibitem{Marcy.1996}
G.~W. {Marcy}, R.~P. {Butler}, {\it \apjl\/} {\bf 464}, L147+ (1996).

\bibitem{Charbonneau.2000}
D.~{Charbonneau}, T.~M. {Brown}, D.~W. {Latham}, M.~{Mayor}, {\it \apjl\/} {\bf
  529}, L45 (2000).

\bibitem{Henry.2000}
G.~W. {Henry}, G.~W. {Marcy}, R.~P. {Butler}, S.~S. {Vogt}, {\it \apjl\/} {\bf
  529}, L41 (2000).

\bibitem{Udalski.2002a}
A.~{Udalski}, {\it et~al.\/}, {\it Acta Astronomica\/} {\bf 52}, 1 (2002).

\bibitem{Udalski.2002b}
A.~{Udalski}, {\it et~al.\/}, {\it Acta Astronomica\/} {\bf 52}, 115 (2002).

\bibitem{Udalski.2003}
A.~{Udalski}, {\it et~al.\/}, {\it Acta Astronomica\/} {\bf 53}, 133 (2003).

\bibitem{McArthur.2004}
B.~E. {McArthur}, {\it et~al.\/}, {\it \apjl\/} {\bf 614}, L81 (2004).

\bibitem{Butler.2004}
P.~{Butler}, {\it et~al.\/}, {\it ArXiv Astrophysics e-prints\/}  (2004).

\bibitem{Narayan.2004}
R.~{Narayan}, A.~{Cumming}, D.~N.~C. {Lin}, {\it ArXiv Astrophysics e-prints\/}
   (2004).

\bibitem{Brown.1936}
E.~W. {Brown}, {\it \mnras\/} {\bf 97}, 62 (1936).

\bibitem{Soderhjelm.1975}
S.~{Soderhjelm}, {\it \aap\/} {\bf 42}, 229 (1975).

\bibitem{Bodenheimer.2003}
P.~{Bodenheimer}, G.~{Laughlin}, D.~N.~C. {Lin}, {\it \apj\/} {\bf 592}, 555
  (2003).

\bibitem{Stoer.1980}
J.~{Stoer}, R.~{Bulirsch}, {\it {Introduction to Numerical Analysis}\/} (New
  York: Springer-Verlag, 1980).

\bibitem{Quinn.1991}
T.~R. {Quinn}, S.~{Tremaine}, M.~{Duncan}, {\it \aj\/} {\bf 101}, 2287 (1991).

\bibitem{Hilditch.2001}
R.~W. {Hilditch}, {\it {An Introduction to Close Binary Stars}\/} (An
  Introduction to Close Binary Stars, by R.W.~Hilditch.~ Cambridge University
  Press, 2001, 392 pp., 2001).

\bibitem{Murray.1999}
C.~D. {Murray}, S.~F. {Dermott}, {\it {Solar System Dynamics}\/} (Cambridge:
  University Press, |c1999, 1999).

\bibitem{Wisdom.1980}
J.~{Wisdom}, {\it \aj\/} {\bf 85}, 1122 (1980).

\bibitem{Laughlin.2001}
G.~{Laughlin}, J.~E. {Chambers}, {\it \apjl\/} {\bf 551}, L109 (2001).

\bibitem{transitsearch}
G.~{Laughlin} (2004). {http://www.transitsearch.org}.

\bibitem{Chambers.1998}
J.~E. {Chambers}, G.~W. {Wetherill}, {\it Icarus\/} {\bf 136}, 304 (1998).

\bibitem{Kokubo.1998}
E.~{Kokubo}, S.~{Ida}, {\it Icarus\/} {\bf 131}, 171 (1998).

\bibitem{Kasting.1993}
J.~F. {Kasting}, D.~P. {Whitmire}, R.~T. {Reynolds}, {\it Icarus\/} {\bf 101},
  108 (1993).

\bibitem{Miralda.2002}
J.~{Miralda-Escud{\' e}}, {\it \apj\/} {\bf 564}, 1019 (2002).

\bibitem{Sasselov.2003}
D.~D. {Sasselov}, {\it \apj\/} {\bf 596}, 1327 (2003).

\end{thebibliography}


\begin{scilastnote}
\item This work was supported in part by the National Science Foundation
under Grant No. PHY99-07949 and by NASA grant NAG5-9678.
This research was supported by NSERC of Canada, and by the Canada
Research Chair program. We are grateful to the Kavli Institute for
Theoretical Physics, where much of this investigation was carried out.
We thank Scott Gaudi and Joshua Winn for helpful discussions and
careful reviews of the manuscript.
\end{scilastnote}

\newpage


\noindent {\bf Fig. 1.} Transit times of the terrestrial planets: The variation of the interval between
successive transits of terrestrial planets, induced by the other
planets in the solar system.  To guide the eye, the solid line connects the times of
each transit.

\noindent {\bf Fig. 2.} Transit times of HDf~209458b: The interval between 
successive transit centers of HD~209458b as a function of time, with
each panel showing the results for a different set of orbital
parameters for the hypothetical second planet. 

\noindent {\bf Fig. 3.} Variations induced by a Jupiter-mass planet: Variations in the interval between successive
transits of a planet with $P_1=3~\mathrm{d}$, $e_1=0.01$, $M_1 = 10^{-3}M_\odot$, induced by a
second planet with mass $M_2 = 10^{-3} M_\odot$. \label{fig:HDintervals3}

\noindent {\bf Fig. 4.} Variations induced by an Earth-mass planet: Variations in the interval between successive
transits of a planet with $P_1=3~\mathrm{d}$, $e_1=0.01$, $M_1 = 10^{-3}M_\odot$, induced by a
second planet with mass $M_2 = 3\times 10^{-6} M_\odot$. \label{fig:HDintervals4}

\begin{figure}
\epsfig{file=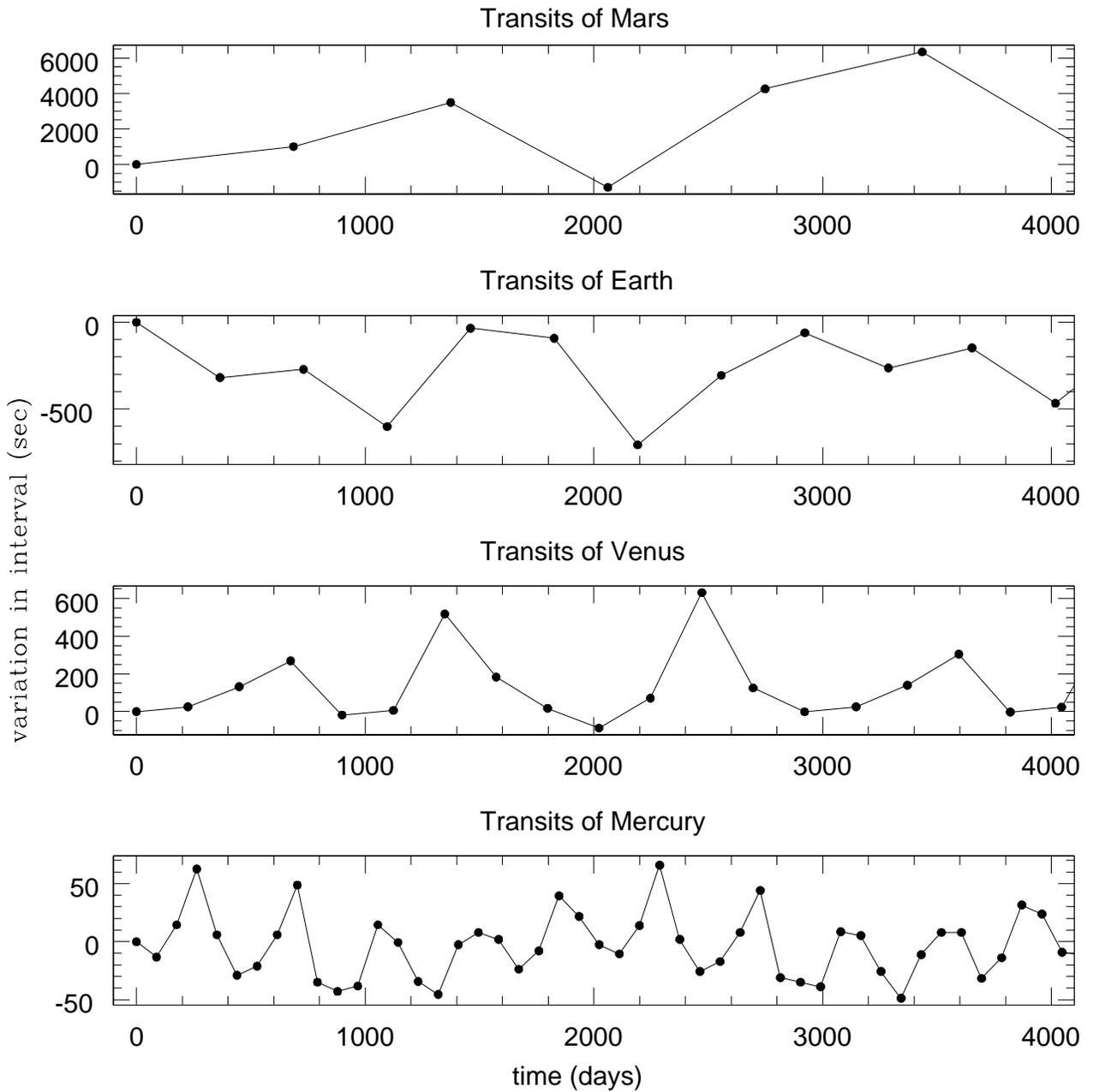, width=7in,height=7in} 
\caption{Variation of transit periods of the terrestrial planets, as
would be recorded by distant observers located in the orbital plane of each planet.}
\label{fig:solar_system}
\end{figure}

\begin{figure}
\epsfig{file=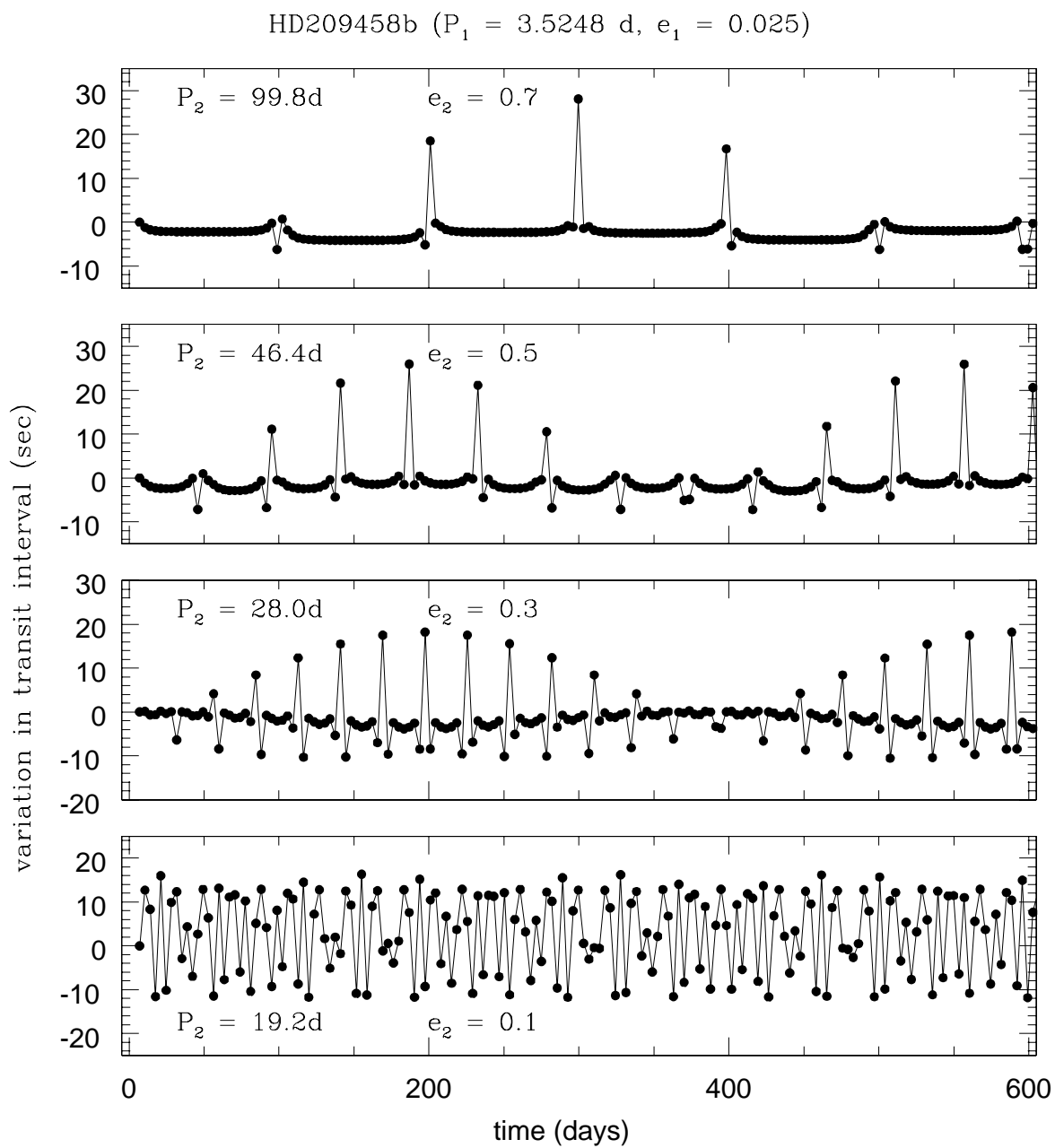, width=7in,height=7in} 
\caption{
Variations of the transit period of HD~209458b, induced by a
hypothetical second planet.}
\label{fig:examples}
\end{figure}

\begin{figure}
\epsfig{file=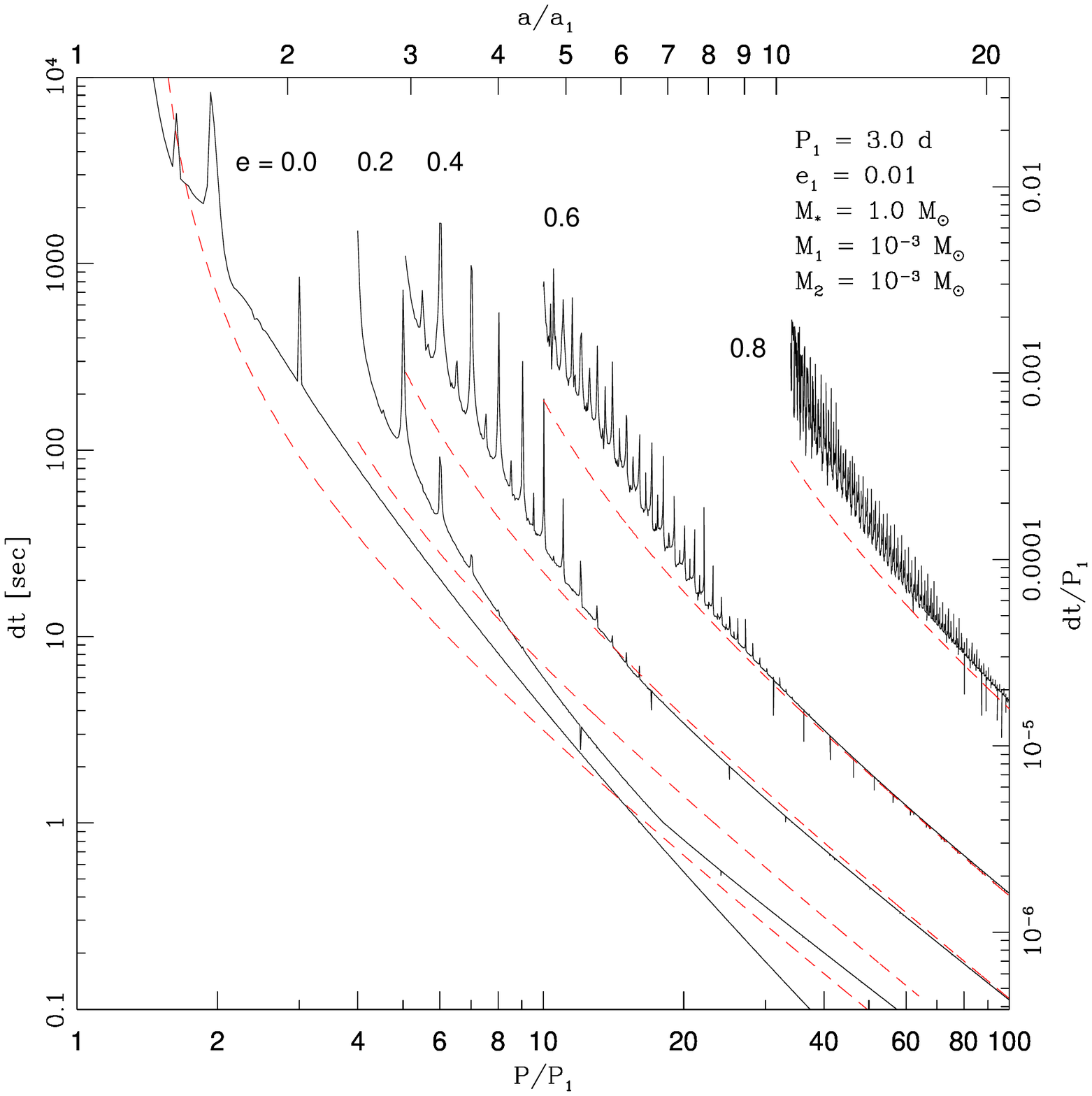, width=7in,height=7in} 
\caption{
Variations in the interval between successive
transits of a planet with $P_1=3~\mathrm{d}$, $e_1=0.01$, $M_1 = 10^{-3}M_\odot$ induced by a
second planet with mass $M_2 = 10^{-3} M_\odot$.}
\label{fig:intervals3}
\end{figure}

\begin{figure}
\epsfig{file=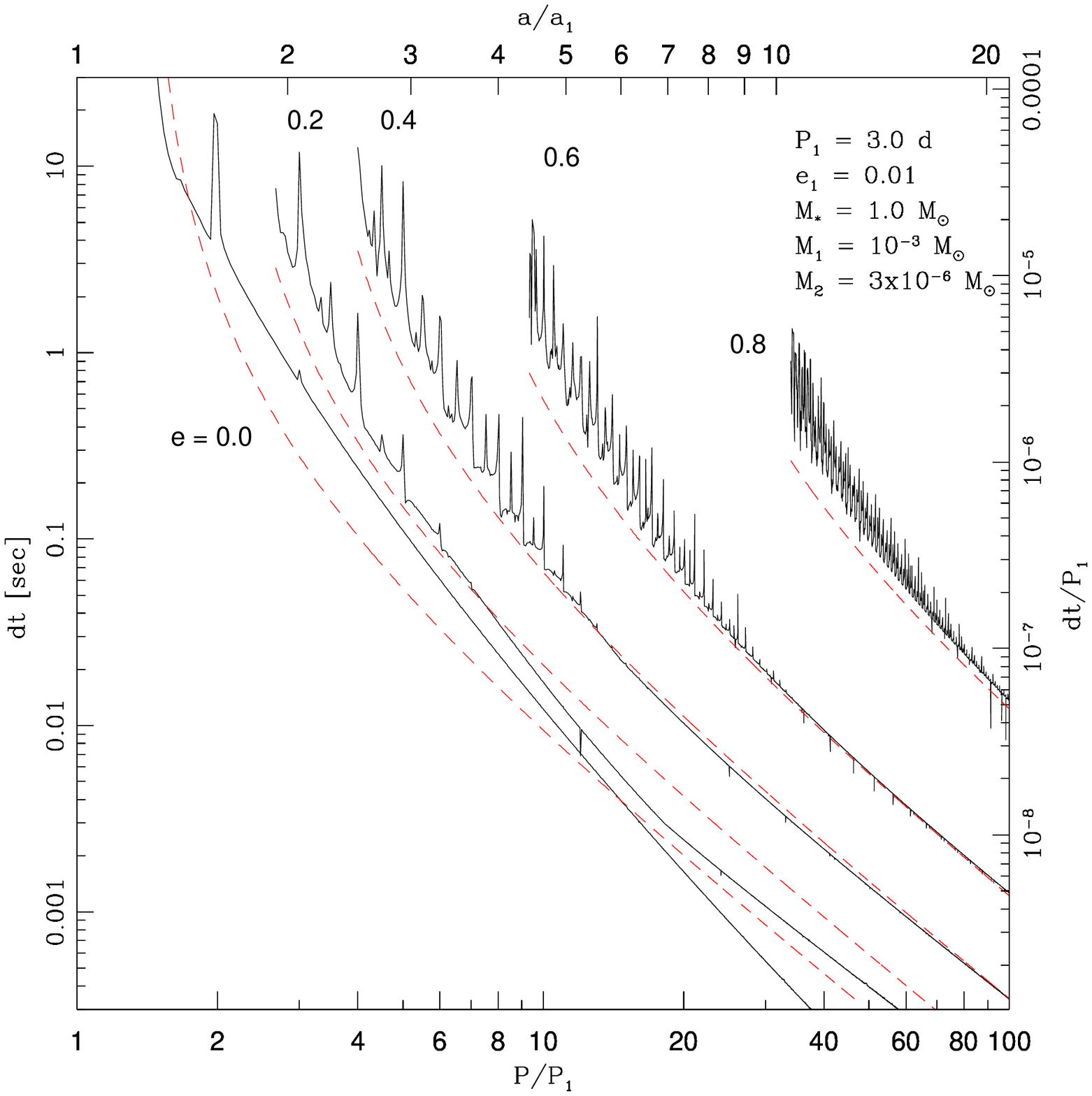, width=7in,height=7in} 
\caption{
Variations in the interval between successive
transits of a planet with $P_1=3~\mathrm{d}$, $e_1=0.01$, $M_1 = 10^{-3}M_\odot$ induced by a
second planet with mass $M_2 = 3\times 10^{-6} M_\odot$.}
\label{fig:intervals4}
\end{figure}

\end{document}